%% file: castex.tex
\documentclass{article}
\usepackage[utf8]{inputenc}
\usepackage[T1]{fontenc}
\usepackage{amsmath}
\usepackage{amsfonts}
\usepackage{latexsym}
\usepackage{amssymb}
\usepackage{times}
\usepackage{ifpdf}
\usepackage{makeidx}
\usepackage{graphicx}
\ifpdf
 \usepackage[pdftex,colorlinks]{hyperref}
\else
 \usepackage[ps2pdf,breaklinks=true,colorlinks=true,linkcolor=red,citecolor=green]{hyperref}
 \fi

\input{giac.tex}
\giacmathjax
\title{Compiling \LaTeX\ to computer algebra-enabled HTML5}
\author{Bernard Parisse\\Institut Fourier\\UMR 5582 du
  CNRS\\Universit\'e de Grenoble I}
\date{2017}
\makeindex
\usepackage{graphicx}
\usepackage{xcolor}
\newcommand{\MarqueCommandeGiac}[1]{
\color[HTML]{8B7500}$\rightarrow$}

\begin{document}
\begin{giacjshere}
\maketitle

\begin{abstract}
This document explains how to create or modify an
existing \LaTeX\ document with commands enabling computations in 
the HTML5 output: when the reader opens the HTML5
output, he can run a computation in his browser, or
modify the command to be executed and run it.
This is done by combining different softwares: \verb|hevea|\cite{hevea}
for compilation to HTML5, \verb|giac.js| for the CAS computing kernel
(itself compiled from the C++ Giac\cite{giac} library with 
\verb|emscripten|\cite{emscripten}), and 
a modified version\cite{hevea2mml} of \verb|itex2MML|\cite{itex2mml} 
for fast and nice rendering in
MathML in browsers that support MathML.
\end{abstract}

\tableofcontents
\printindex

\section{Introduction}
Combining \LaTeX\ rendering quality and CAS computing is not new:
\begin{enumerate}
\item math softwares provide converters to export data to a \LaTeX\ file,
or provide automated computations in a way similar 
to the way bibtex provides bibliography, like sagetex (\cite{sagetex}).
\item some softwares handle both \LaTeX-like rendering and
computation, for example texmacs (\cite{texmacs}), lyx (\cite{lyx}), 
Jupyter notebook (\cite{jupyter}). 
 \end{enumerate}
However, in the first case,
the reader can not modify the CAS commandlines, and in the second
case the data format is not standard \LaTeX\ (the writer can not
start from an existing document)
and requires additional software to be installed on
the reader device or a net access to a server 
to run the computations.

The solution presented here is new in that the writer will edit
a standard \LaTeX\ file, add a few easy to learn commands like
\verb|\giacinputmath{factor(x^10-1)}| or \verb|\giacinput{plot(sin(x))}|
and compile it to produce a HTML5+MathML
document. 
The reader can see the document in any browser 
(it's optimized for Firefox), without installation, and he can 
modify computation commandlines
and run them on his own computer.

If you are reading this file in PDF format, it is highly recommended to
open 
\footahref{https://www-fourier.ujf-grenoble.fr/\home{parisse}/giac/castex.html}{the HTML5/Mathml version}
in order to test interactivity
and look at the \LaTeX\ 
\footahref{https://www-fourier.ujf-grenoble.fr/\home{parisse}/giac/castex.tex}{source}

\section{User manual}
\subsection{Installation on the writer computer}
The writer must install 
\begin{itemize}
\item the latest unstable version
of \footahref{http://hevea.inria.fr/distri/unstable/}{{\tt hevea}} 
(\cite{hevea})
or a forked version 
\footahref{https://github.com/YannickChevalier/hevea-mathjax}{{\tt hevea-mathjax}} (\cite{heveamathjax})),
\item \footahref{https://www-fourier.ujf-grenoble.fr/\home{parisse}/giac.html}{{\tt Giac/Xcas}} (\cite{giac}) 
for computing-enabled output
\item \footahref{https://www-fourier.ujf-grenoble.fr/\home{parisse}/giac/heveatomml.tgz}{\tt heveatomml} (\cite{hevea2mml}) 
for MathML output
\end{itemize}
The files 
\footahref{https://www-fourier.ujf-grenoble.fr/\home{parisse}/giac/giac.tex}{{\tt giac.tex}} (or the French version
\footahref{https://www-fourier.ujf-grenoble.fr/\home{parisse}/giac/giacfr.tex}{{\tt giacfr.tex}})
\footahref{https://www-fourier.ujf-grenoble.fr/\home{parisse}/giac.js}{{\tt giac.js}}, 
\footahref{https://www-fourier.ujf-grenoble.fr/\home{parisse}/giac/hevea.sty}{{\tt hevea.sty}}, 
\footahref{https://www-fourier.ujf-grenoble.fr/\home{parisse}/giac/mathjax.sty}{{\tt mathjax.sty}} must
be copied in the \LaTeX\ working directory.
On an Internet connected linux box, the writer can run once
the following shell script to install\index{install} the tools required 
for HTML5/MathML output~:
\begin{verbatim}
#! /bin/bash
wget https://www-fourier.ujf-grenoble.fr/~parisse/giac/giac.tex
wget https://www-fourier.ujf-grenoble.fr/~parisse/giac/giacfr.tex
wget https://www-fourier.ujf-grenoble.fr/~parisse/giac/giac.js
wget https://www-fourier.ujf-grenoble.fr/~parisse/giac/hevea.sty
wget https://www-fourier.ujf-grenoble.fr/~parisse/giac/mathjax.sty
wget http://hevea.inria.fr/distri/unstable/hevea-2017-05-18.tar.gz
tar xvfz hevea-2017-05-18.tar.gz
cd hevea-2017-05-18
make
sudo make install
cd ..
wget https://www-fourier.ujf-grenoble.fr/~parisse/giac/heveatomml.tgz
tar xvfz heveatomml.tgz
cd heveatomml/src
make
sudo make install
cd ../..
\end{verbatim}

\subsection{On the writer side}
We now assume that the installation is done.
The writer opens a \LaTeX\ file with his usual editor. He must add in
the preamble the following lines
\begin{verbatim}
\makeindex
\input{giac.tex}

\giacmathjax
\end{verbatim}
 For interactive CAS \LaTeX\ commands support,
the writer should add 
\begin{verbatim}
\begin{giacjshere}
\tableofcontents
\printindex
\end{verbatim}
just after \verb|\begin{document}| and 
\begin{verbatim}
\end{giacjshere}
\end{verbatim}
just before \verb|\end{document}|.
Printing the table of contents and index before the first \LaTeX\
section command is recommended, otherwise the HTML output
\verb|Table| and \verb|Index| buttons will not link correctly.

The rest of the source file is standard \LaTeX\ except that
\begin{itemize}
\item References to numbered equations should be inside additional 
backslash-ed parenthesis, for example
\ifhevea
\begin{rawhtml}
\begin{verbatim}
\begin{equation} \label{eq_test}
 \frac{2}{x^2-1}=\frac{1}{x-1}-\frac{1}{x+1} 
\end{equation}
From equation (\(\ref{eq_test}\)) ...
\end{verbatim}
\end{rawhtml}
\else
\begin{verbatim}
\begin{equation} \label{eq:test}
 \frac{2}{x^2-1}=\frac{1}{x-1}-\frac{1}{x+1} 
\end{equation}
From equation (\(\ref{eq:test}\)) ...
\end{verbatim}
\fi
\begin{equation} \label{eq:test}
 \frac{2}{x^2-1}=\frac{1}{x-1}-\frac{1}{x+1} 
\end{equation}
From equation (\(\ref{eq:test}\)) ...
\item \verb|\mathbb{}| should be explicit, commands like
\verb|\R| where \verb|\R| is defined
by \verb|\newcommand{\R}{\mathbb{R}}| will not work.
\item New commands are available for interactive CAS support
\begin{itemize}
\item 
\verb|\giacinputmath{commandline}|\index{giacinputmath} will output 
an inline
\verb|commandline| 
that the user can modify and execute, the answer will be displayed
in MathML (or SVG for 2-d graph output).\\
Example~: \verb|\giacinputmath{factor(x^10-1)}|\\

\verb|factor(x^10-1)|\\
$$(x-1)\cdot (x+1) (x^{4}-x^{3}+x^{2}-x+1) (x^{4}+x^{3}+x^{2}+x+1)$$
\\
{\bf Warnings}, if your command contains \verb|<| or \verb|>|, you must
replace them by \verb|&lt;| or \verb|&gt;|, 
otherwise they will be interpreted as HTML delimiters. You can also
use the \verb|giacprog| and \verb|giaconload| environments explained below.\\
If the output 
is a 2-d graph, do not skip a line with \verb|\\| after the command 
for PDF output
\item
\verb|\giaccmdmath{command}{arguments}|\index{giaccmdmath} 
will output \verb|command| in a button following the \verb|arguments|, 
the reader can only modify the arguments:\\
\verb|\giaccmdmath{factor}{x^4-1}|\index{giaccmdmath}\\

\verb|factor(x^4-1)|\\
$$(x-1)\cdot (x+1) (x^{2}+1)$$
\\
\item 
These commands may take an optional HTML style argument, for example\\
\verb|\giacinputmath[style="width:200px;"]{factor(x^10-1)}|\\

\verb|factor(x^10-1)|\\
$$(x-1)\cdot (x+1) (x^{4}-x^{3}+x^{2}-x+1) (x^{4}+x^{3}+x^{2}+x+1)$$
\\
\verb|\giaccmdmath[style="font-size:x-large"]{factor}{x^4-1}|\\

\verb|factor(x^4-1)|\\
$$(x-1)\cdot (x+1) (x^{2}+1)$$
\\
\item
There are similar commands for outlined output
\verb|\giacinputbigmath{}|\index{giacinputbigmath} 
or \verb|\giaccmdbigmath{}{}|\index{giaccmdbigmath}:\\
For example \verb|\giacinputbigmath{factor(x^25-1)}|\\

\verb|factor(x^25-1)|\\
$$(x-1) (x^{4}+x^{3}+x^{2}+x+1) (x^{20}+x^{15}+x^{10}+x^{5}+1)$$
\\
Example with an optional style argument 
\verb|\giacinputbigmath[style="width:600px;height:20px;"]{factor(x^25-1)}|\\

\verb|factor(x^25-1)|\\
$$(x-1) (x^{4}+x^{3}+x^{2}+x+1) (x^{20}+x^{15}+x^{10}+x^{5}+1)$$
\\
\verb|\giaccmdbigmath{factor}{x^25-1}|\index{giaccmdbigmath}\\

\verb|factor(x^25-1)|\\
$$(x-1) (x^{4}+x^{3}+x^{2}+x+1) (x^{20}+x^{15}+x^{10}+x^{5}+1)$$
\\
\verb|\giaccmdbigmath[style="width:600px;height:20px;"]{factor}{x^25-1}|\\

\verb|factor(x^25-1)|\\
$$(x-1) (x^{4}+x^{3}+x^{2}+x+1) (x^{20}+x^{15}+x^{10}+x^{5}+1)$$
\item Similar commands with text (or plot) output 
\verb|\giacinput|\index{giacinput} and \verb|\giacinputbig|\index{giacinputbig} and
\verb|\giaccmd|\index{giaccmd}, 
example:\\
\verb|\giacinput{factor(x^4-1)}|\index{giacinput}~:\\

\verb|factor(x^4-1)|\\
$$(x-1)\cdot (x+1) (x^{2}+1)$$
\\
\verb|\giaccmd{print}{"Hello world"}|\index{giaccmd}~:\\

\verb|print("Hello world")|\\
$$0$$
\\
With optional style argument\\
\verb|\giacinput[style="font-size:x-large"]{plot(1/x)}|\\

\verb|plot(1/x)|\\

\begin{center}
\includegraphics[width=0.8\linewidth]{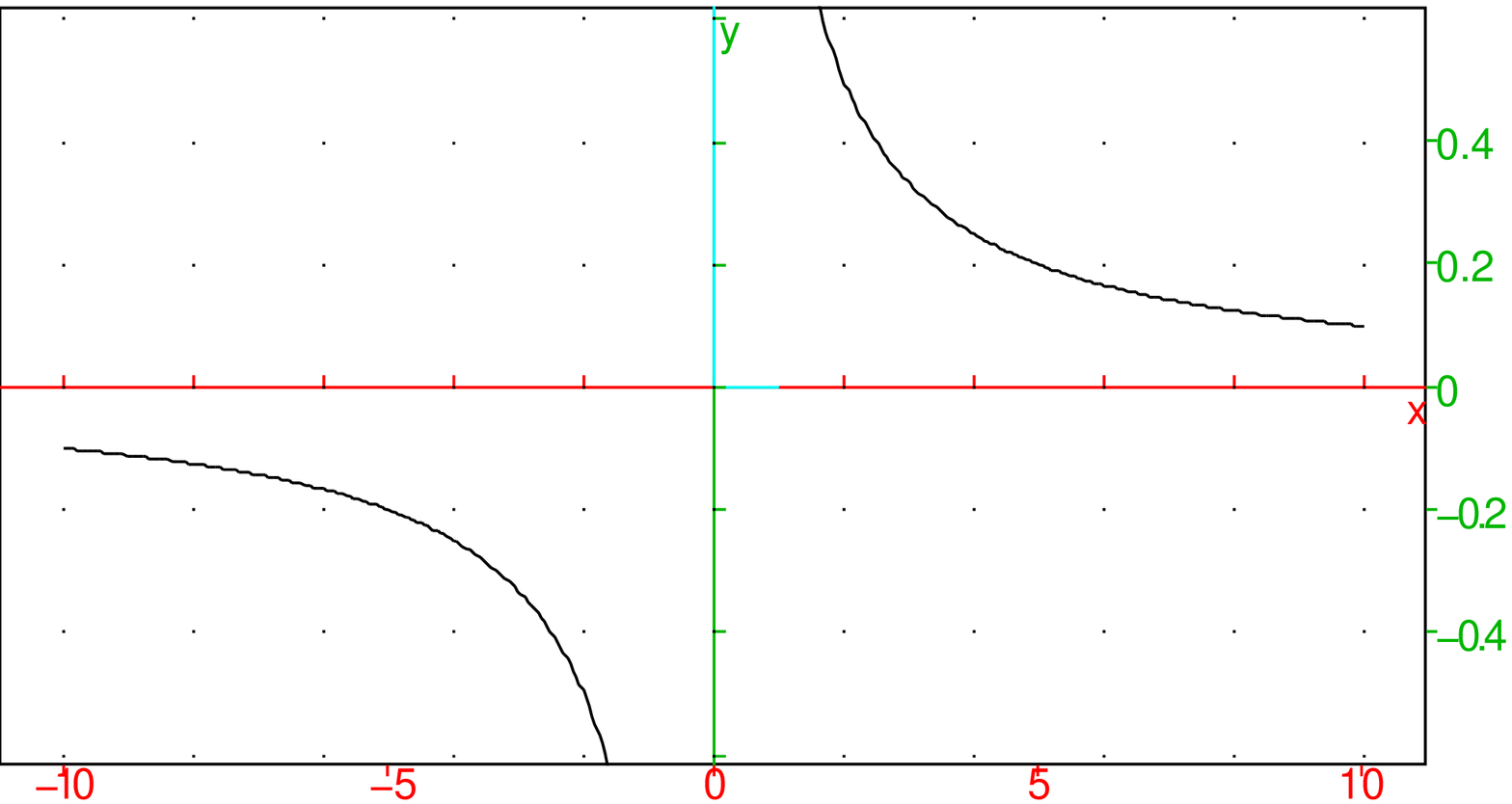}
\end{center}

\verb|\giaccmd[style="font-size:x-large"]{factor}{x^4-1}|\\

\verb|factor(x^4-1)|\\
$$(x-1)\cdot (x+1) (x^{2}+1)$$
\\

\item
The \verb|giacprog|\index{giacprog} environment should be used for programs or multi-line commands\\
\verb|\begin{giacprog}...\end{giacprog}|\\
Inside this environment, you can keep \verb|<| and \verb|>|. The
program will be parsed once the user press the \verb|ok| button. After
parse, the program may be modified and parsed again.\\
{\bf Warning}: Do not use the \verb|giacprog| environment in another environment (like itemize or
enumerate).\\
If you want the program to be parsed at load-time, replace
\verb|giacprog| with \verb|giaconload|\index{giaconload}:\\
\verb|\begin{giaconload}...\end{giaconload}|
\item
The \verb|\giacslider{idname}{min}{max}{step}{value}{command}| 
command will add
a slider. When the user modifies the slider interactively, the new
\verb|value| is stored in \verb|idname| and the \verb|command| 
(depending on \verb|idname|)
is executed. Example:\\ 
\verb|\giacslider{a}{-5}{5}{0.1}{0.5}{plot(sin(a*x))}|\index{giacslider}\\

\verb|a:=0.5;plot(sin(a*x))|\\

\begin{center}
\includegraphics[width=0.8\linewidth]{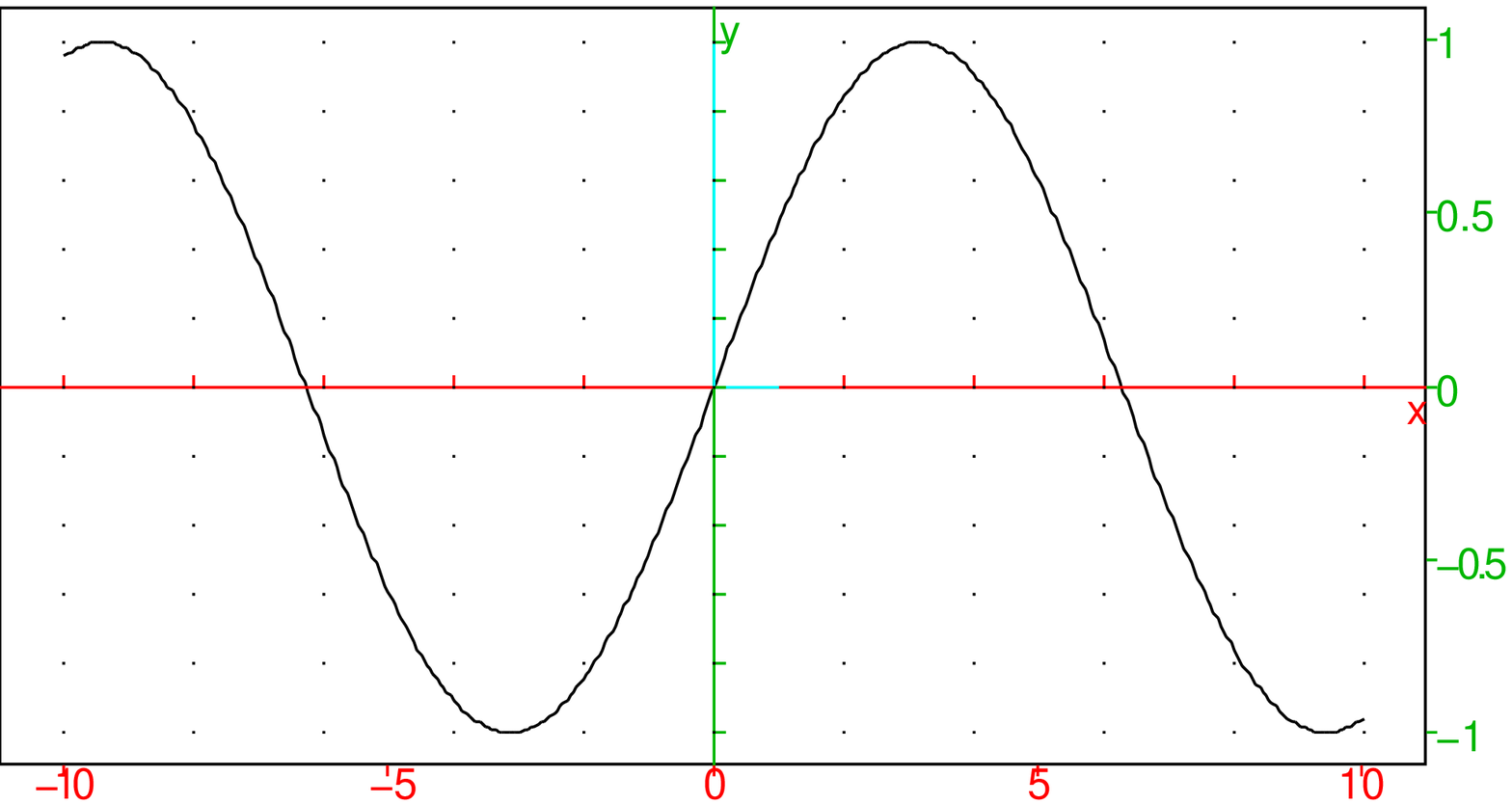}
\end{center}

\item
The \verb|\giachidden|\index{giachidden} command behaves like
\verb|\giaccmd| except that the default HTML5 style is ``hidden''
until the command button has been pressed.

\item
The \verb|\giaclink|\index{giaclink} command will add a link in the
HTML version and nothing in PDF/DVI. The links open in a new tab, and the
corresponding text may be specified as optional argument (default 
is Test online). Note that \verb|hevea.sty| provides similar commands
(\verb|\ahref|, \verb|\footahref|, \verb|\ahrefurl|) 
with output in PDF/DVI.\\
Example with a link to Xcas for Firefox with a few commands\\
\verb|\giaclink{http://www-fourier.ujf-grenoble.fr/\%7eparisse/xcasen.html#+factor(x^4-1)&+a:=idn(3)&}|\\

\giaclink{http://www-fourier.ujf-grenoble.fr/\%7eparisse/xcasen.html#+factor(x^4-1)&+a:=idn(3)&}
\end{itemize}
\end{itemize}

Once the source file is written, it is compiled to HTML5 with the command\\
\verb|hevea2mml sourcefile.tex|\\
The HTML output and the \verb|giac.js| files should be in the same
directory on the web server.
Index and bibliography should be processed with \verb|makeindex|
and \verb|bibhva|.

If a PDF output is desired,
the command \verb|icas| from a Giac/Xcas installation
should be used
instead of \verb|pdflatex| 
because it will run all CAS commands, output them in a temporary
\LaTeX\ file, 
and run \verb|pdflatex| on the output 
(this was inspired by the 
\footahref{http://melusine.eu.org/syracuse/giac/pgiac/}{pgiac script} 
from Jean-Michel Sarlat \cite{pgiac}). The temporary file name
is obtained by adding a {\tt\_ } at the end of the initial file name
(without the {\tt .tex} extension). Therefore, if you have 
an index and or citations, you should run \verb|makeindex| and \verb|bibtex|
on the file name with \verb|_| appended. For \verb|bibtex| citations in
the HTML files, you should run \verb|bibhva|.
For example, the PDF version of this document is available
\footahref{https://www-fourier.ujf-grenoble.fr/\home{parisse}/giac/castex.pdf}{here}.

\subsection{On the reader side}
The reader's browser opens an HTML5+MathML file (linking to the JavaScript
\verb|giac.js|). The MathML is rendered natively on Firefox or Safari,
while Chrome or Internet Explorer will automatically load MathJax to
render MathML (this is of course noticeably slower if the document is
large).
Computations are run by the reader's browser (the CAS is JavaScript
code). This is slower than native code but faster than net access to
a server and it does not require setting up a specific server for
computations.

\subsection{More examples}
\subsubsection{Trace (2-d graph)}
This example illustrates with a slider that the evolute of a curve 
is the envelope of the normals to the curve, here the curve is an ellipsis
and the envelop an astroid. 
The list of normals \verb|L| is initialized empty at load-time.
\begin{verbatim}
L:=[]\end{verbatim}
Now move the slider:

\begin{verbatim}
t0:=0.7;gl_x=-6..6;gl_y=-4..4;G:=plotparam
([2*cos(t),sin(t)],t=0..2*pi);M:=element
(G,evalf(t0));T:=tangent(M);N:=perpendicular
(M,T);L:=append(L,N);evolute(G,color=red)\end{verbatim}

\begin{center}
\includegraphics[width=0.8\linewidth]{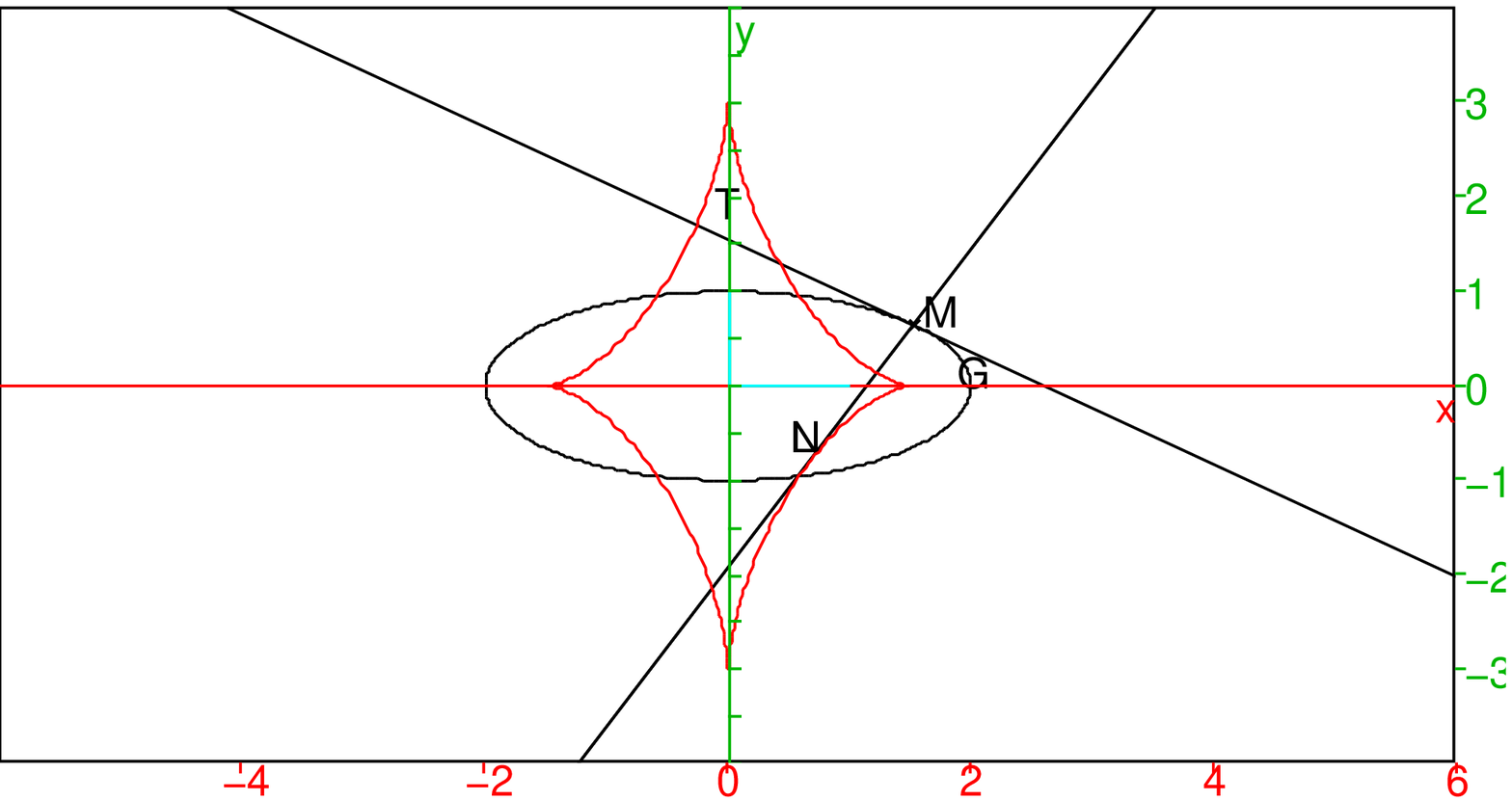}
\end{center}

\subsubsection{Cone section (3-d graph)}
$C$ is a cone of center the origin, axis of direction $(0,0,1)$, and angle
$\frac{\pi}{6}$, $P$ is a plane of equation $z=my+3$. 
$m$ is controlled by the slider, when $m$ moves the intersection 
is an ellipsis or hyperbola (limit value is a parabola).

\begin{verbatim}
m:=0.7;C:=cone([0,0,0],[0,0,1],pi/6, display=green+filled
);P:=plane(z=evalf(m)*y+3,display=cyan+filled);\end{verbatim}

\begin{center}
\includegraphics[width=0.8\linewidth]{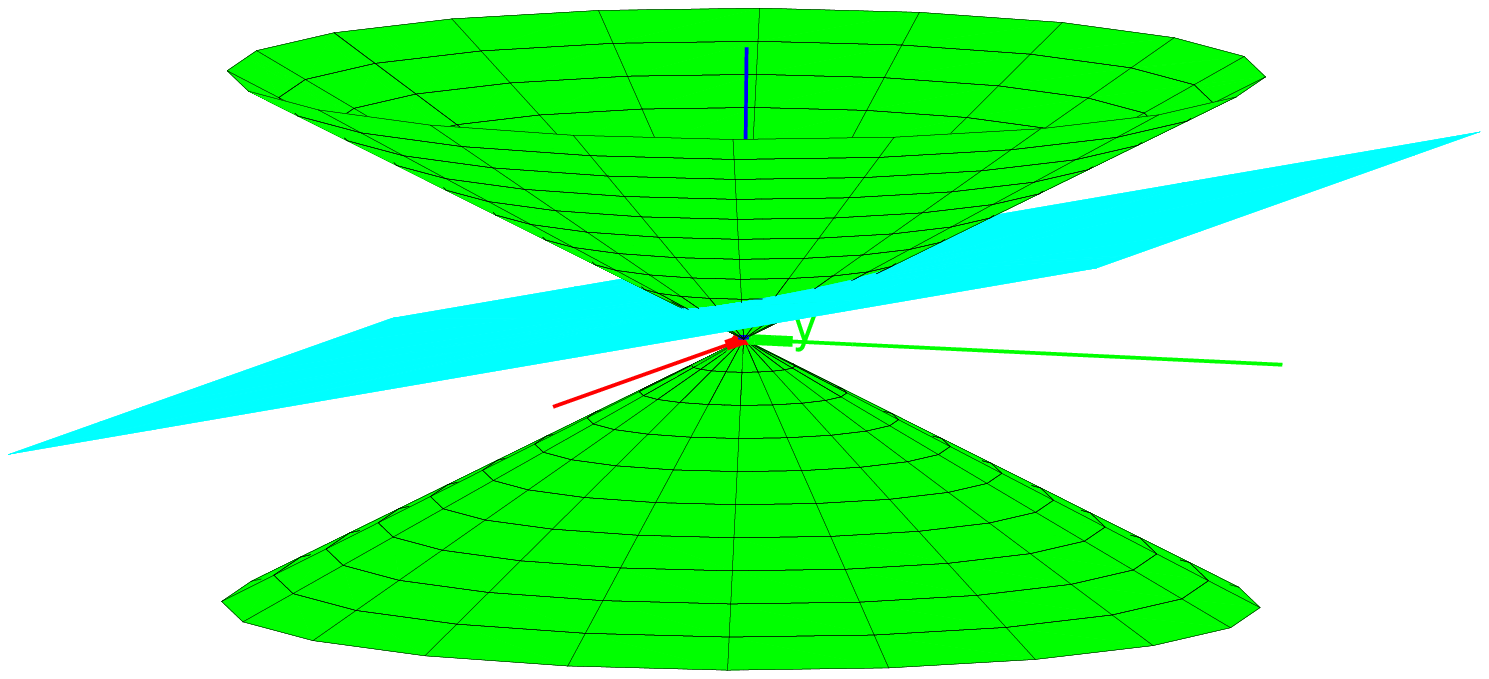}
\end{center}

\subsubsection{Dunford decomposition (CAS)}
A program computing the Dunford decomposition
of a matrix with Newton method. 
It is parsed at load-time (\verb|giaconload| environment).
\begin{verbatim}
function dunford(A)
  local U,p,q,q1,j,d,n;
  U:=A;
  n:=nrows(U);
  p:=charpoly(U);
  q:=p/gcd(p,p'); // square free part
  q1:=q';
  for (j:=1;j<=n;j:=2*j){
    d:=inv(horner(q1,U))*horner(q,U); // Newton step
    if (d==0*d) return U,A-U;
    U:=U-d;
  }
  return U,A-U;
end:;
\end{verbatim}
Example~: we define $J$ an almost diagonal matrix and $A$ a similar matrix
and we check the Dunford decomposition of $A$.
$$ J=\left(\begin{array}{ccc}
2 & 0 & 0 \\
0 & 1 & 1 \\
0 & 0 & 1
\end{array}\right) , \quad
P=\left(\begin{array}{ccc}
1 & 0 & 0 \\
2 & -1 & 0 \\
3 & 4 & 1
\end{array}\right), \quad
A=PJP^{-1} $$

\begin{verbatim}
J:=[[2,0,0],[0,1,1],[0,0,1]]; P:=trn([[1,2,3
],[0,-1,4],[0,0,1]]); A:=P*J*inv(P)\end{verbatim}
$$\left(\begin{array}{ccc}
2 & 0 & 0 \\
0 & 1 & 1 \\
0 & 0 & 1
\end{array}\right) ,\left(\begin{array}{ccc}
1 & 0 & 0 \\
2 & -1 & 0 \\
3 & 4 & 1
\end{array}\right) ,\left(\begin{array}{ccc}
2 & 0 & 0 \\
13 & -3 & -1 \\
-41 & 16 & 5
\end{array}\right) $$
\\

\verb|D,N:=dunford(A); N^2; P*diag(diag(J))*inv(P)|\\
$$\left(\begin{array}{ccc}
2 & 0 & 0 \\
2 & 1 & 0 \\
3 & 0 & 1
\end{array}\right) ,\left(\begin{array}{ccc}
0 & 0 & 0 \\
11 & -4 & -1 \\
-44 & 16 & 4
\end{array}\right) ,\left(\begin{array}{ccc}
0 & 0 & 0 \\
0 & 0 & 0 \\
0 & 0 & 0
\end{array}\right) ,\left(\begin{array}{ccc}
2 & 0 & 0 \\
2 & 1 & 0 \\
3 & 0 & 1
\end{array}\right) $$
\\

\subsubsection{Slopefield}
This will display the slopefield of an ordinary differential equation
$$\frac{dy}{dt}=-y+cos(t)$$
and one solution corresponding to an initial condition $y(0)$ that the user may
modify with the slider.

\begin{verbatim}
y0:=1.0;gl_x=-5..5; gl_y=-3..3;plotfield
(-y+cos(t),[t=-5..5,y=-3..3],xstep=0.4,ystep=0.4
);plotode(-y+cos(t),[t=-5..5,y],[0,y0],tstep=0.1,color=red
)\end{verbatim}

\begin{center}
\includegraphics[width=0.8\linewidth]{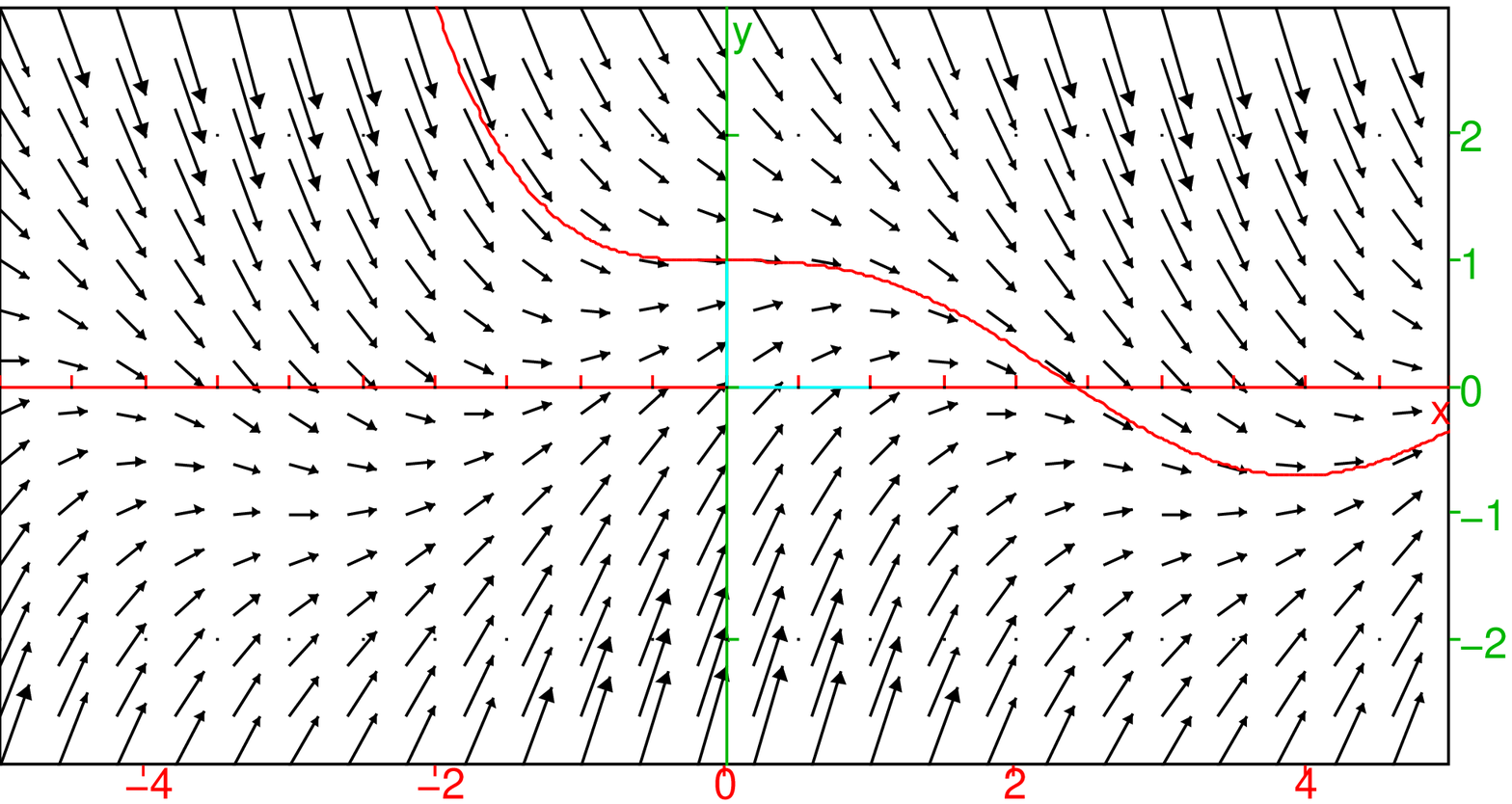}
\end{center}

\subsubsection{Gr\"obner basis (CAS)}
The CAS kernel can compute non-trivial Gr\"obner basis. Of course,
the JavaScript version is significantly slower than the native Giac/Xcas
kernel.
\begin{verbatim}
kat7:=[-x1+2*x8^2+2*x7^2+2*x6^2+2*x5^2+2*x4^2+2*x3^2+2*x2^2+x1^2,
 -x2+2*x8*x7+2*x7*x6+2*x6*x5+2*x5*x4+2*x4*x3+2*x3*x2+2*x2*x1,
 -x3+2*x8*x6+2*x7*x5+2*x6*x4+2*x5*x3+2*x4*x2+2*x3*x1+x2^2,
 -x4+2*x8*x5+2*x7*x4+2*x6*x3+2*x5*x2+2*x4*x1+2*x3*x2,
 -x5+2*x8*x4+2*x7*x3+2*x6*x2+2*x5*x1+2*x4*x2+x3^2,
 -x6+2*x8*x3+2*x7*x2+2*x6*x1+2*x5*x2+2*x4*x3,
 -x7+2*x8*x2+2*x7*x1+2*x6*x2+2*x5*x3+x4^2,
 -1+2*x8+2*x7+2*x6+2*x5+2*x4+2*x3+2*x2+x1]:;
\end{verbatim}
Basis over $\mathbb{Z}/16777213$

\begin{verbatim}
G:=gbasis(kat7 mod 16777213,[x1,x2,x3,x4,x5,x6,x7,x8
]):; size(G); G[20];\end{verbatim}
$$\mathrm{Done},74,\mathrm{x5}\cdot \mathrm{x7}\cdot \mathrm{x8}^{4}+6710886\cdot \mathrm{x4}\cdot \mathrm{x8}^{5}-3938997\cdot \mathrm{x5}\cdot \mathrm{x8}^{5}+5106109\cdot \mathrm{x6}\cdot \mathrm{x8}^{5}-5543774\cdot \mathrm{x7}\cdot \mathrm{x8}^{5}+4960220\cdot \mathrm{x8}^{6}-1622886\cdot \mathrm{x7}^{5}+3350574\cdot \mathrm{x7}^{4}\cdot \mathrm{x8}-3203410\cdot \mathrm{x7}^{3}\cdot \mathrm{x8}^{2}-6856774\cdot \mathrm{x2}\cdot \mathrm{x6}\cdot \mathrm{x8}^{3}-2852328\cdot \mathrm{x3}\cdot \mathrm{x6}\cdot \mathrm{x8}^{3}+4830623\cdot \mathrm{x4}\cdot \mathrm{x6}\cdot \mathrm{x8}^{3}+6654447\cdot \mathrm{x2}\cdot \mathrm{x7}\cdot \mathrm{x8}^{3}+1485963\cdot \mathrm{x3}\cdot \mathrm{x7}\cdot \mathrm{x8}^{3}+7645914\cdot \mathrm{x4}\cdot \mathrm{x7}\cdot \mathrm{x8}^{3}+6108299\cdot \mathrm{x5}\cdot \mathrm{x7}\cdot \mathrm{x8}^{3}-5645530\cdot \mathrm{x7}^{2}\cdot \mathrm{x8}^{3}+1467630\cdot \mathrm{x2}\cdot \mathrm{x8}^{4}-4480913\cdot \mathrm{x3}\cdot \mathrm{x8}^{4}-6905666\cdot \mathrm{x4}\cdot \mathrm{x8}^{4}-7864873\cdot \mathrm{x5}\cdot \mathrm{x8}^{4}+5036573\cdot \mathrm{x6}\cdot \mathrm{x8}^{4}-3080880\cdot \mathrm{x7}\cdot \mathrm{x8}^{4}-4359461\cdot \mathrm{x8}^{5}-4003610\cdot \mathrm{x2}\cdot \mathrm{x7}^{3}-4722776\cdot \mathrm{x3}\cdot \mathrm{x7}^{3}-1804635\cdot \mathrm{x4}\cdot \mathrm{x7}^{3}-161589\cdot \mathrm{x5}\cdot \mathrm{x7}^{3}+5344773\cdot \mathrm{x6}\cdot \mathrm{x7}^{3}+8336020\cdot \mathrm{x7}^{4}-3781288\cdot \mathrm{x2}\cdot \mathrm{x4}\cdot \mathrm{x7}\cdot \mathrm{x8}-3554207\cdot \mathrm{x2}\cdot \mathrm{x5}\cdot \mathrm{x7}\cdot \mathrm{x8}+5772217\cdot \mathrm{x3}\cdot \mathrm{x5}\cdot \mathrm{x7}\cdot \mathrm{x8}-1758721\cdot \mathrm{x2}\cdot \mathrm{x7}^{2}\cdot \mathrm{x8}+6818342\cdot \mathrm{x3}\cdot \mathrm{x7}^{2}\cdot \mathrm{x8}-2283117\cdot \mathrm{x4}\cdot \mathrm{x7}^{2}\cdot \mathrm{x8}+2932624\cdot \mathrm{x5}\cdot \mathrm{x7}^{2}\cdot \mathrm{x8}+8254803\cdot \mathrm{x6}\cdot \mathrm{x7}^{2}\cdot \mathrm{x8}-5242202\cdot \mathrm{x7}^{3}\cdot \mathrm{x8}+8076837\cdot \mathrm{x2}\cdot \mathrm{x4}\cdot \mathrm{x8}^{2}+17492\cdot \mathrm{x2}\cdot \mathrm{x5}\cdot \mathrm{x8}^{2}+2985896\cdot \mathrm{x3}\cdot \mathrm{x5}\cdot \mathrm{x8}^{2}-7321877\cdot \mathrm{x2}\cdot \mathrm{x6}\cdot \mathrm{x8}^{2}+6542761\cdot \mathrm{x3}\cdot \mathrm{x6}\cdot \mathrm{x8}^{2}+1168714\cdot \mathrm{x4}\cdot \mathrm{x6}\cdot \mathrm{x8}^{2}+7247300\cdot \mathrm{x2}\cdot \mathrm{x7}\cdot \mathrm{x8}^{2}-6008550\cdot \mathrm{x3}\cdot \mathrm{x7}\cdot \mathrm{x8}^{2}-2288236\cdot \mathrm{x4}\cdot \mathrm{x7}\cdot \mathrm{x8}^{2}+3659549\cdot \mathrm{x5}\cdot \mathrm{x7}\cdot \mathrm{x8}^{2}+3059392\cdot \mathrm{x6}\cdot \mathrm{x7}\cdot \mathrm{x8}^{2}+7854288\cdot \mathrm{x7}^{2}\cdot \mathrm{x8}^{2}+4699527\cdot \mathrm{x2}\cdot \mathrm{x8}^{3}+2770983\cdot \mathrm{x3}\cdot \mathrm{x8}^{3}-361720\cdot \mathrm{x4}\cdot \mathrm{x8}^{3}-1732766\cdot \mathrm{x5}\cdot \mathrm{x8}^{3}+7556960\cdot \mathrm{x6}\cdot \mathrm{x8}^{3}+4537622\cdot \mathrm{x7}\cdot \mathrm{x8}^{3}+6304804\cdot \mathrm{x8}^{4}+3315662\cdot \mathrm{x2}\cdot \mathrm{x4}\cdot \mathrm{x6}+4416381\cdot \mathrm{x6}^{3}+8014147\cdot \mathrm{x2}\cdot \mathrm{x4}\cdot \mathrm{x7}+6313405\cdot \mathrm{x2}\cdot \mathrm{x5}\cdot \mathrm{x7}-6166461\cdot \mathrm{x3}\cdot \mathrm{x5}\cdot \mathrm{x7}-5677397\cdot \mathrm{x2}\cdot \mathrm{x6}\cdot \mathrm{x7}+5361345\cdot \mathrm{x3}\cdot \mathrm{x6}\cdot \mathrm{x7}-1317626\cdot \mathrm{x4}\cdot \mathrm{x6}\cdot \mathrm{x7}+4866646\cdot \mathrm{x6}^{2}\cdot \mathrm{x7}-983362\cdot \mathrm{x2}\cdot \mathrm{x7}^{2}-4165951\cdot \mathrm{x3}\cdot \mathrm{x7}^{2}+7598477\cdot \mathrm{x4}\cdot \mathrm{x7}^{2}+2841104\cdot \mathrm{x5}\cdot \mathrm{x7}^{2}+1175912\cdot \mathrm{x6}\cdot \mathrm{x7}^{2}+256538\cdot \mathrm{x7}^{3}+5505000\cdot \mathrm{x2}\cdot \mathrm{x4}\cdot \mathrm{x8}-797220\cdot \mathrm{x2}\cdot \mathrm{x5}\cdot \mathrm{x8}+712446\cdot \mathrm{x3}\cdot \mathrm{x5}\cdot \mathrm{x8}+4754075\cdot \mathrm{x2}\cdot \mathrm{x6}\cdot \mathrm{x8}-2132948\cdot \mathrm{x3}\cdot \mathrm{x6}\cdot \mathrm{x8}+2532426\cdot \mathrm{x4}\cdot \mathrm{x6}\cdot \mathrm{x8}+5979911\cdot \mathrm{x6}^{2}\cdot \mathrm{x8}+5610090\cdot \mathrm{x2}\cdot \mathrm{x7}\cdot \mathrm{x8}+7153809\cdot \mathrm{x3}\cdot \mathrm{x7}\cdot \mathrm{x8}+5595623\cdot \mathrm{x4}\cdot \mathrm{x7}\cdot \mathrm{x8}-1819263\cdot \mathrm{x5}\cdot \mathrm{x7}\cdot \mathrm{x8}-2038671\cdot \mathrm{x6}\cdot \mathrm{x7}\cdot \mathrm{x8}+7135075\cdot \mathrm{x7}^{2}\cdot \mathrm{x8}+1582261\cdot \mathrm{x2}\cdot \mathrm{x8}^{2}+1104606\cdot \mathrm{x3}\cdot \mathrm{x8}^{2}+2300040\cdot \mathrm{x4}\cdot \mathrm{x8}^{2}+6997133\cdot \mathrm{x5}\cdot \mathrm{x8}^{2}-309985\cdot \mathrm{x6}\cdot \mathrm{x8}^{2}+1805598\cdot \mathrm{x7}\cdot \mathrm{x8}^{2}+6256405\cdot \mathrm{x8}^{3}+4736670\cdot \mathrm{x2}\cdot \mathrm{x4}-2786037\cdot \mathrm{x2}\cdot \mathrm{x5}-4067833\cdot \mathrm{x3}\cdot \mathrm{x5}+6347609\cdot \mathrm{x2}\cdot \mathrm{x6}+304173\cdot \mathrm{x3}\cdot \mathrm{x6}-8228932\cdot \mathrm{x4}\cdot \mathrm{x6}+8076963\cdot \mathrm{x5}\cdot \mathrm{x6}-2387968\cdot \mathrm{x6}^{2}-6160826\cdot \mathrm{x2}\cdot \mathrm{x7}-3201521\cdot \mathrm{x3}\cdot \mathrm{x7}-5829581\cdot \mathrm{x4}\cdot \mathrm{x7}+8039746\cdot \mathrm{x5}\cdot \mathrm{x7}-8070323\cdot \mathrm{x6}\cdot \mathrm{x7}+5282236\cdot \mathrm{x7}^{2}+8160369\cdot \mathrm{x2}\cdot \mathrm{x8}+7695372\cdot \mathrm{x3}\cdot \mathrm{x8}+3981391\cdot \mathrm{x4}\cdot \mathrm{x8}+2313354\cdot \mathrm{x5}\cdot \mathrm{x8}+2267196\cdot \mathrm{x6}\cdot \mathrm{x8}+479769\cdot \mathrm{x7}\cdot \mathrm{x8}-7762287\cdot \mathrm{x8}^{2}-6102183\cdot \mathrm{x2}+2233312\cdot \mathrm{x3}+7143090\cdot \mathrm{x4}+1632345\cdot \mathrm{x5}+4462465\cdot \mathrm{x6}+2457609\cdot \mathrm{x7}-4797781\cdot \mathrm{x8}$$
Basis over $\mathbb{Q}$

\begin{verbatim}
G:=gbasis(kat7,[x1,x2,x3,x4,x5,x6,x7,x8]
):; size(G); G[20];\end{verbatim}
$$\mathrm{Done},74,18489624116678107161957583274880000\cdot \mathrm{x5}\cdot \mathrm{x7}\cdot \mathrm{x8}^{4}+14791699293342485729566066619904000\cdot \mathrm{x4}\cdot \mathrm{x8}^{5}+15434816653922593804764591255552000\cdot \mathrm{x5}\cdot \mathrm{x8}^{5}+12862347211602161503970492712960000\cdot \mathrm{x6}\cdot \mathrm{x8}^{5}+13505464572182269579169017348608000\cdot \mathrm{x7}\cdot \mathrm{x8}^{5}+9325201728411567090378607216896000\cdot \mathrm{x8}^{6}-7187105831817036884041483200000\cdot \mathrm{x7}^{5}-33336388571521098989490364800000\cdot \mathrm{x7}^{4}\cdot \mathrm{x8}-105003336171975847577258519232000\cdot \mathrm{x7}^{3}\cdot \mathrm{x8}^{2}+160779340145027018799631158912000\cdot \mathrm{x2}\cdot \mathrm{x6}\cdot \mathrm{x8}^{3}+600221686826000076111666520128000\cdot \mathrm{x3}\cdot \mathrm{x6}\cdot \mathrm{x8}^{3}+1273847681214291488932343388672000\cdot \mathrm{x4}\cdot \mathrm{x6}\cdot \mathrm{x8}^{3}+186437307964613840475659253936000\cdot \mathrm{x2}\cdot \mathrm{x7}\cdot \mathrm{x8}^{3}+752203505074579054604747812972800\cdot \mathrm{x3}\cdot \mathrm{x7}\cdot \mathrm{x8}^{3}-455139490175965453027265051155200\cdot \mathrm{x4}\cdot \mathrm{x7}\cdot \mathrm{x8}^{3}-3732767314807254778553012658144000\cdot \mathrm{x5}\cdot \mathrm{x7}\cdot \mathrm{x8}^{3}+454447652288092171148808976608000\cdot \mathrm{x7}^{2}\cdot \mathrm{x8}^{3}-8262037479986243633645642476800\cdot \mathrm{x2}\cdot \mathrm{x8}^{4}-2123145455568590801717110005264000\cdot \mathrm{x3}\cdot \mathrm{x8}^{4}-6448947272961384722334143186246400\cdot \mathrm{x4}\cdot \mathrm{x8}^{4}-9020080899608501266861863690144000\cdot \mathrm{x5}\cdot \mathrm{x8}^{4}-4592423657176534614634493662368000\cdot \mathrm{x6}\cdot \mathrm{x8}^{4}-4068981625139405428536903185164800\cdot \mathrm{x7}\cdot \mathrm{x8}^{4}-6256804721103899791153992476409600\cdot \mathrm{x8}^{5}-50930003285005805746034112864000\cdot \mathrm{x2}\cdot \mathrm{x7}^{3}-98815694902455856040050013078400\cdot \mathrm{x3}\cdot \mathrm{x7}^{3}-178904425012942489930525712256000\cdot \mathrm{x4}\cdot \mathrm{x7}^{3}-169461108998411771955789499924800\cdot \mathrm{x5}\cdot \mathrm{x7}^{3}-227603798249140831496645498832000\cdot \mathrm{x6}\cdot \mathrm{x7}^{3}-283005930977935211726310113683200\cdot \mathrm{x7}^{4}-71492789590179998478096859200000\cdot \mathrm{x2}\cdot \mathrm{x4}\cdot \mathrm{x7}\cdot \mathrm{x8}-38175240284719596365474879712000\cdot \mathrm{x2}\cdot \mathrm{x5}\cdot \mathrm{x7}\cdot \mathrm{x8}-211622019575626043231720640768000\cdot \mathrm{x3}\cdot \mathrm{x5}\cdot \mathrm{x7}\cdot \mathrm{x8}-12294800535637866450286796313600\cdot \mathrm{x2}\cdot \mathrm{x7}^{2}\cdot \mathrm{x8}-244544205095213710871048306682240\cdot \mathrm{x3}\cdot \mathrm{x7}^{2}\cdot \mathrm{x8}-514369866356213144883063355497600\cdot \mathrm{x4}\cdot \mathrm{x7}^{2}\cdot \mathrm{x8}-1130209578595706816454261839304480\cdot \mathrm{x5}\cdot \mathrm{x7}^{2}\cdot \mathrm{x8}-1055393444465211695689618263691200\cdot \mathrm{x6}\cdot \mathrm{x7}^{2}\cdot \mathrm{x8}-1406119502686437528035756838448320\cdot \mathrm{x7}^{3}\cdot \mathrm{x8}-23700715356981790571308637572800\cdot \mathrm{x2}\cdot \mathrm{x4}\cdot \mathrm{x8}^{2}-224307271739787199648124398358400\cdot \mathrm{x2}\cdot \mathrm{x5}\cdot \mathrm{x8}^{2}-703117776418774938586467121219200\cdot \mathrm{x3}\cdot \mathrm{x5}\cdot \mathrm{x8}^{2}-668883607291375835229514134691200\cdot \mathrm{x2}\cdot \mathrm{x6}\cdot \mathrm{x8}^{2}-1416129544171117132305479191140480\cdot \mathrm{x3}\cdot \mathrm{x6}\cdot \mathrm{x8}^{2}-1693309491669927189427175709302400\cdot \mathrm{x4}\cdot \mathrm{x6}\cdot \mathrm{x8}^{2}-395265349264428254843128402464960\cdot \mathrm{x2}\cdot \mathrm{x7}\cdot \mathrm{x8}^{2}-2157829953471193803435578880936000\cdot \mathrm{x3}\cdot \mathrm{x7}\cdot \mathrm{x8}^{2}-2287246775616834619187840512512240\cdot \mathrm{x4}\cdot \mathrm{x7}\cdot \mathrm{x8}^{2}-2139437629827332642047433869035360\cdot \mathrm{x5}\cdot \mathrm{x7}\cdot \mathrm{x8}^{2}-2497871812000497177519762356428560\cdot \mathrm{x6}\cdot \mathrm{x7}\cdot \mathrm{x8}^{2}-3492608318155396470100069650530880\cdot \mathrm{x7}^{2}\cdot \mathrm{x8}^{2}-1039486205141588203865998989816960\cdot \mathrm{x2}\cdot \mathrm{x8}^{3}-520881635849620160061396194144400\cdot \mathrm{x3}\cdot \mathrm{x8}^{3}-583118783545795304031042474015360\cdot \mathrm{x4}\cdot \mathrm{x8}^{3}-306723722820160083550375066483680\cdot \mathrm{x5}\cdot \mathrm{x8}^{3}-1050610603648552085949448750151040\cdot \mathrm{x6}\cdot \mathrm{x8}^{3}-2953141371900339833402393044080240\cdot \mathrm{x7}\cdot \mathrm{x8}^{3}+374371816008112991685996200902080\cdot \mathrm{x8}^{4}-1141506944295511933372122489600\cdot \mathrm{x2}\cdot \mathrm{x4}\cdot \mathrm{x6}+271109377228003452783896549760\cdot \mathrm{x6}^{3}+2628441194615113394423827720800\cdot \mathrm{x2}\cdot \mathrm{x4}\cdot \mathrm{x7}+9613536060917368128768219000000\cdot \mathrm{x2}\cdot \mathrm{x5}\cdot \mathrm{x7}+12192741404432093251997475825600\cdot \mathrm{x3}\cdot \mathrm{x5}\cdot \mathrm{x7}+5564895255194648753981939738400\cdot \mathrm{x2}\cdot \mathrm{x6}\cdot \mathrm{x7}-10893156890484284402862821580180\cdot \mathrm{x3}\cdot \mathrm{x6}\cdot \mathrm{x7}-34648546550014310221447329206520\cdot \mathrm{x4}\cdot \mathrm{x6}\cdot \mathrm{x7}+5188507391232533844112778323560\cdot \mathrm{x6}^{2}\cdot \mathrm{x7}+7632836308610801201148972855300\cdot \mathrm{x2}\cdot \mathrm{x7}^{2}-14188559176224136293959154885120\cdot \mathrm{x3}\cdot \mathrm{x7}^{2}+3871815792450784509518290458180\cdot \mathrm{x4}\cdot \mathrm{x7}^{2}+89868955193762132338520852933280\cdot \mathrm{x5}\cdot \mathrm{x7}^{2}+118738983748114730979236687763600\cdot \mathrm{x6}\cdot \mathrm{x7}^{2}+112700208596061313641314862430560\cdot \mathrm{x7}^{3}+4053544256098076583899196741600\cdot \mathrm{x2}\cdot \mathrm{x4}\cdot \mathrm{x8}+6865996284226614668583179489760\cdot \mathrm{x2}\cdot \mathrm{x5}\cdot \mathrm{x8}+45242337648974041470614272158060\cdot \mathrm{x3}\cdot \mathrm{x5}\cdot \mathrm{x8}+32227201282706955721618206220800\cdot \mathrm{x2}\cdot \mathrm{x6}\cdot \mathrm{x8}+74541032189966350904022813664440\cdot \mathrm{x3}\cdot \mathrm{x6}\cdot \mathrm{x8}+109222371672216817132192001508000\cdot \mathrm{x4}\cdot \mathrm{x6}\cdot \mathrm{x8}-17611004640614086324388992411080\cdot \mathrm{x6}^{2}\cdot \mathrm{x8}+29594299342945246138164657151080\cdot \mathrm{x2}\cdot \mathrm{x7}\cdot \mathrm{x8}+205159980647763425790697815649620\cdot \mathrm{x3}\cdot \mathrm{x7}\cdot \mathrm{x8}+278339206993654959199265198747760\cdot \mathrm{x4}\cdot \mathrm{x7}\cdot \mathrm{x8}+320109418092634463592305998443900\cdot \mathrm{x5}\cdot \mathrm{x7}\cdot \mathrm{x8}+470455163686395205853079522680640\cdot \mathrm{x6}\cdot \mathrm{x7}\cdot \mathrm{x8}+755778577030191710047502491424040\cdot \mathrm{x7}^{2}\cdot \mathrm{x8}+154034703044709254756761011059520\cdot \mathrm{x2}\cdot \mathrm{x8}^{2}+408438843078288837589395409099080\cdot \mathrm{x3}\cdot \mathrm{x8}^{2}+626571774184586400445578730691040\cdot \mathrm{x4}\cdot \mathrm{x8}^{2}+624320625021505001235797755627800\cdot \mathrm{x5}\cdot \mathrm{x8}^{2}+765354057100611716543860177215120\cdot \mathrm{x6}\cdot \mathrm{x8}^{2}+1613656634668411336764387754269240\cdot \mathrm{x7}\cdot \mathrm{x8}^{2}+742017541276853107569936500714280\cdot \mathrm{x8}^{3}-569978404721446095488940896160\cdot \mathrm{x2}\cdot \mathrm{x4}+249618277062914161062989489490\cdot \mathrm{x2}\cdot \mathrm{x5}+1900410268192910546526379348920\cdot \mathrm{x3}\cdot \mathrm{x5}+4235302367045310263654245214340\cdot \mathrm{x2}\cdot \mathrm{x6}+6017966457428446868635215553710\cdot \mathrm{x3}\cdot \mathrm{x6}+7191909101684961100934533192860\cdot \mathrm{x4}\cdot \mathrm{x6}+9789374444237912565456468704848\cdot \mathrm{x5}\cdot \mathrm{x6}+11559501634033175787315078653112\cdot \mathrm{x6}^{2}+101079410207415260494159169730\cdot \mathrm{x2}\cdot \mathrm{x7}+99661074822647098112134792800\cdot \mathrm{x3}\cdot \mathrm{x7}-109112809723521688004909929982\cdot \mathrm{x4}\cdot \mathrm{x7}-1338027341456674595447033958096\cdot \mathrm{x5}\cdot \mathrm{x7}+13966570096620920111910735849858\cdot \mathrm{x6}\cdot \mathrm{x7}+6167950709721386348534611467920\cdot \mathrm{x7}^{2}+6479267803700614230481232223780\cdot \mathrm{x2}\cdot \mathrm{x8}-7414048412549736404950116605942\cdot \mathrm{x3}\cdot \mathrm{x8}-24353719861736871295970330985036\cdot \mathrm{x4}\cdot \mathrm{x8}-39701066286451965695455145707452\cdot \mathrm{x5}\cdot \mathrm{x8}-61550928718514540066256221335760\cdot \mathrm{x6}\cdot \mathrm{x8}-114900330861316836907381280674920\cdot \mathrm{x7}\cdot \mathrm{x8}-181315500248057737735069888116216\cdot \mathrm{x8}^{2}+145410814833489474708302735228\cdot \mathrm{x2}+567461515246797856334772898352\cdot \mathrm{x3}+690332069209059893002349525292\cdot \mathrm{x4}+810398183902583651863724586368\cdot \mathrm{x5}+1317646827183272993428371499025\cdot \mathrm{x6}+2648636588012522741659902630708\cdot \mathrm{x7}+2995670767262379465274814083712\cdot \mathrm{x8}$$

\section{How this is done}
The \LaTeX\ \verb|\giac...| commands are defined in \verb|giac.tex|.
For example \verb|\giacinput| is defined like this:
\begin{verbatim}
\newcommand{
\verb||\\
$$\,\mathrm{undef}\,$$
[2][style="width:400px;font-size:large"]{
\ifhevea
\@print{<textarea onkeypress="UI.ckenter(event,this,1)" }
\@getprint{#1>#2}
\@print{</textarea><button onclick="previousSibling.style.display='inherit';var tmp=UI.caseval(previousSibling.value);tmp=UI.rmquote(tmp); nextSibling.innerHTML='&nbsp;'+tmp;UI.render_canvas(nextSibling);">ok</button><span></span><br>}
\else
\lstinline@#2@
\fi
}
\end{verbatim}
If \verb|hevea| compiles the command, the \verb|\ifhevea| part is
active, and the command will output an HTML5 \verb|<textarea>| element
and a OK \verb|<button>|, with a callback to JavaScript code
that will evaluate the CAS command inside the textarea \\
\verb|var tmp=UI.caseval(previousSibling.value)|\\
and fill the next HTML5 \verb|<span>| field with the result of the CAS
command.

The CAS evaluation is performed by a call to \verb|giaceval| in the
\verb|UI.caseval| code (defined in \verb|giac.tex|), where
\verb|giaceval| is a global JavaScript variable assigned at page load-time
from the \verb|Module| interface created by compiling Giac/Xcas with
the C++ to JavaScript compiler 
\footahref{http://kripken.github.io/emscripten-site/}{{\tt emscripten}}. 
The CAS code being in JavaScript, it can be run on every
JavaScript-enabled browser. It will be faster on browsers that have
support for \verb|asm.js| (\verb|asmjs.org|) 
like Mozilla Firefox: numerical computations
are 1 to 2 times slower than native code, while exact computations
are 2 to 10 times slower than native code (the main reason being that
JavaScript has currently no 64 bits integer type).

For a PDF output, if \verb|pdflatex| is run on the tex file,
giac commands will be written verbatim, but they will not be processed.
The \verb|icas| command from the Giac/Xcas package will filter all
giac commands, process them and output the result in math mode in a
temporary \LaTeX\ file. If the
answer is a 2-d graph output, \verb|icas| will output a pdf file on
the hard disk and output a corresponding \verb|\includegraphics|
command in the temporary \LaTeX\ file. After that, the temporary file 
will be processed by \verb|pdflatex|.

\section{Conclusion}
The current version of \verb|icas| and \verb|giac.tex| are already
usable to easily produce HTML interactive CAS-enabled document from 
\LaTeX\ documents. They may be completed in future versions depending
on user requests. For example, online courses might have commands to
enable student exercises answers auto-check.

{\bf Acknowledgements} \\
Thanks to Luc Maranget and Yannick Chevalier for fixing bugs in
mathjax-enabled hevea. 
Thanks to Ren\'ee De Graeve and Murielle Stepec who have tested
preliminary versions of this compilation method.

\bibliography{latex.bib}

\end{giacjshere}
\end{document}

%% file: giac.tex
\usepackage{hevea} 
\usepackage{listings}
\usepackage{fancyvrb}
\ifhevea 
\newcommand\giacmathjax{
\usepackage[auto]{mathjax}
\renewcommand{\jax@meta}{\begin{rawhtml}<script language="javascript"> 
var ua = window.navigator.userAgent;  
var old_ie = ua.indexOf('MSIE ');  
var new_ie = ua.indexOf('Trident/');  
if ((old_ie > -1) || (new_ie > -1) || Boolean(window.chrome)){
 (function () {
  var script = document.createElement("script");
  script.type = "text/javascript";
  script.src  = "https://cdnjs.cloudflare.com/ajax/libs/mathjax/2.7.0/MathJax.js?config=TeX-MML-AM_CHTML";
  document.getElementsByTagName("head")[0].appendChild(script);
})();
}
</script>
\end{rawhtml}}
}
\newenvironment{giacjs}[1]["max-height: 500px; overflow:auto"]
{\loadgiacmain{#1}
}
{\loadgiaccontrol
\loadgiacscriptstart
\@print{
<script src="file:///usr/share/giac/doc/giac.js" async></script> 
}
\loadgiacscriptend
} 
\newenvironment{giacjshere}[1]["max-height: 500px; overflow:auto"]
{\loadgiacmain{#1}
}
{\loadgiaccontrol
\loadgiacscriptstart
\@print{
<script src="giac.js" async></script> 
}
\loadgiacscriptend
} 
\newenvironment{giacparijs}[1]["max-height: 500px; overflow:auto"]
{\loadgiacmain{#1}
}
{
\loadgiaccontrol
\loadgiacscriptstart
\@print{
<script src="file:///usr/share/giac/doc/giacpari.js" async></script> 
}
\loadgiacscriptend
} 
\newenvironment{giacjsonline}[1]["max-height: 500px; overflow:auto"]
{\loadgiacmain{#1}
}
{
\loadgiaccontrol
\loadgiacscriptstart
\@print{
<script src="https://www-fourier.ujf-grenoble.fr/~parisse/giac.js" async></script> 
}
\loadgiacscriptend
} 
\newenvironment{giacparijsonline}[1]["max-height: 500px; overflow:auto"]
{\loadgiacmain{#1}
}
{
\loadgiaccontrol
\loadgiacscriptstart
\@print{
<script src="https://www-fourier.ujf-grenoble.fr/~parisse/giacpari.js" async></script> 
}
\loadgiacscriptend
} 
\else
\newcommand\giacmathjax{}

\newenvironment{giacjshere}[1]["max-height: 500px; overflow:auto"]{}{}

\fi
\newcommand{\loadgiacscriptstart}{
\ifhevea
\@print{
<script language="javascript"> 
var Module = { 
        htmlcheck:true,
        htmlbuffer:'',
        preRun: [],
        postRun: [],
        print: (function() {
          var element = document.getElementById('output');
          element.innerHTML='';// element.value = ''; // clear browser cache
          return function(text) {
            //console.log(text.charCodeAt(0));
            if (text.length==1 && text.charCodeAt(0)==12){ element.innerHTML=''; return; }
            if (text.length>=1 && text.charCodeAt(0)==2) {console.log('STX');Module.htmlcheck=false; htmlbuffer='';return;}
            if (text.length>=1 && text.charCodeAt(0)==3) {console.log('ETX');Module.htmlcheck=true; element.style.display='inherit'; element.innerHTML += htmlbuffer;htmlbuffer='';element.scrollTop = 99999; return;}
            if (Module.htmlcheck){
            // These replacements are necessary if you render to raw HTML 
             text = text.replace(/&/g, "&amp;");
             text = text.replace(/</g, "&lt;");
             text = text.replace(/>/g, "&gt;");
             text = text.replace('\n', '<br>', 'g');
             text += '<br>'
             element.style.display='inherit';
             element.innerHTML += text; // element.value += text + "\n";
             element.scrollTop = 99999; // focus on bottom
            } else htmlbuffer += text;
          };
        })(),
     canvas: document.getElementById('canvas'),};
</script>
}
\fi
}
\newcommand{\loadgiacmain}[1]{
\ifhevea
\@print{
<div>
<div id="maindiv" style=}#1\@print{>}
\fi
}
\newcommand{\loadgiaccontrol}{
\ifhevea
\@print{
</div>
<div id="controldiv" style="max-height: 400px; overflow:auto">
<span id="controlindex"></span>
<button title="Clone last command to Xcas on line" onclick="UI.clone()">Clone</button>
<button title="Clear console" onclick="document.getElementById('output').innerHTML='';">Clear</button>
<button title="Increase console size" onclick="var field=document.getElementById('output'); var s=field.style.maxHeight; s=s.substr(0,s.length-2);s=eval(s)+20 ;if(s<innerHeight/2){s=s+'px';field.style.maxHeight=s; field=document.getElementById('maindiv'); s=field.style.maxHeight; s=s.substr(0,s.length-2);s=eval(s)-20; s=s+'px';field.style.maxHeight=s;}">+</button>
<button title="Decrease console size" onclick="var field=document.getElementById('output'); var s=field.style.maxHeight; s=s.substr(0,s.length-2);s=eval(s)-20 ; if(s>80){s=s+'px';field.style.maxHeight =s;field=document.getElementById('maindiv'); s=field.style.maxHeight; s=s.substr(0,s.length-2);s=eval(s)+20; s=s+'px';field.style.maxHeight=s;}">-</button>
<textarea title="Commandline for a quick computation" onkeypress="UI.ckenter(event,this,3);" style="width:400px;height:20px;font-size:large"></textarea><button  title="Eval previous cell" onclick="UI.quick(this);">-></button><span></span>
&nbsp;&nbsp;<button onclick="var s=UI.caseval('restart;'); Module.print(s);" title="Reset CAS computing kernel">Restart</button>
<button onclick="UI.exec(document.documentElement);" title='Click here to exec all commands (maybe be long!)'>Exec. all</button>
<canvas id='canvas' width=0 height=0   onmousedown="UI.canvas_pushed=true;UI.canvas_lastx=event.clientX; UI.canvas_lasty=event.clientY;"  onmouseup="UI.canvas_pushed=false;" onmousemove="UI.canvas_mousemove(event,'')"></canvas>
<div id="output" style="max-height: 80px; overflow:auto"></div>
</div>
</div>}
\fi
}
\ifhevea
\newcommand{\loadgiacscriptend}{
\@print{
<script language="javascript"> 
 var UI = {
  histcount:0,
  usemathjax:false,
  lastcmd:'',
  clone:function(){
    if (UI.lastcmd.length){
       var tmp=giaceval('VARS(-1)');
       tmp=encodeURIComponent(tmp);
       var url='https://www-fourier.ujf-grenoble.fr/~parisse/xcasen.html#+'+tmp+':;&+'+UI.lastcmd;
       console.log(url);
       window.open(url,'_blank');
    }
  },
  canvas_pushed:false,
  canvas_lastx:0,
  canvas_lasty:0,
  canvas_mousemove:function(event,no){
    if (UI.canvas_pushed){
      // Module.print(event.clientX);
      if (UI.canvas_lastx!=event.clientX){
        if (event.clientX>UI.canvas_lastx)
          giac3d('r'+no);
        else
          giac3d('l'+no);
        UI.canvas_lastx=event.clientX;
      }
      if (UI.canvas_lasty!=event.clientY){
        if (event.clientY>UI.canvas_lasty)
          giac3d('d'+no);
        else
          giac3d('u'+no);
        UI.canvas_lasty=event.clientY;
      }
    }
  },  
  render_canvas:function(field){
   var n=field.id;
   if (n && n.length>5 && n.substr(0,5)=='gl3d_'){
    Module.print(n);
    var n3d=n.substr(5,n.length-5);
    giac3d(n3d);
    return;
   }
   var f=field.firstChild;
   for (;f;f=f.nextSibling){
     UI.render_canvas(f);
   }
  },
  count_newline:function(s){
    var ss=s.length,i,res=1;
    for (i=0;i<ss;i++){
      if (s[i]=='\n') res++;
    }
    return res;
  },
  ltgt:function(s){
    var ss=s.length,i,res='',c=0;
    for (i=0;i<ss-4;i++){
      if (s[i]=='\n') c++;      
      if (s[i]!='&' || s[i+2]!='t' || s[i+3]!=';'){
        res += s[i];
        continue;
      }
      if (s[i+1]=='l'){
        res += '<';
        i +=3;
        continue;
      }
      if (s[i+1]=='g'){
        res += '>';
        i +=3;
        continue;
      }
      res += s[i];
    }
    for (;i<ss;i++) res+=s[i];
    return [res,c];
  },
  rmquote:function(tmp){
    var s=tmp.length;
    if (s>2 && tmp.charCodeAt(0)==34 && tmp.charCodeAt(s-1)==34)
      tmp=tmp.substr(1,s-2);
    return tmp;
  },
  quick:function(field){
    var tmp1=field.previousSibling.value;
    var tmp=UI.caseval(tmp1);
    tmp=UI.rmquote(tmp); 
    tmp=UI.latexeval(tmp);
    Module.print(String.fromCharCode(2));
    Module.print("<tt>");
    Module.print(tmp1);
    Module.print("</tt><br>&nbsp;&nbsp;");
    Module.print(tmp);
    Module.print("<br>");
    Module.print(String.fromCharCode(3));
    field.nextSibling.innerHTML='&nbsp;'+tmp;
   //UI.render_canvas(nextSibling);  
  },
  caseval:function(text){
    UI.lastcmd=text;
    var s="not evaled",err;
    try {
       s= giaceval(text);
    } catch (err) { s=err.message;}
    var is_3d=s.length>5 && s.substr(0,5)=='gl3d ';
    if (is_3d){
	var n3d=s.substr(5,s.length-5);
	s = '<canvas id="gl3d_'+n3d+'" onmousedown="UI.canvas_pushed=true;UI.canvas_lastx=event.clientX; UI.canvas_lasty=event.clientY;" onmouseup="UI.canvas_pushed=false;" onmousemove="UI.canvas_mousemove(event,'+n3d+')" width=400 height=250></canvas>';
    }
   //console.log(s);
    return s;
  },
  eval_form: function(field){
    giaceval('assume('+field.name.value+'='+field.valname.value+')');
    var s=UI.caseval(field.prog.value);
    var is_svg=s.substr(1,4)=='<svg';
    if (is_svg) field.parentNode.lastChild.innerHTML=s.substr(1,s.length-2);
    else field.parentNode.lastChild.innerHTML=s;
   UI.render_canvas(field.parentNode.lastChild);
  },
  latexeval:function(text){
    var tmp=text;
    if (tmp.length>5 && tmp.substr(0,5)=='gl3d_') return tmp;
    if (tmp.length>5 && tmp.substr(1,4)=='<svg') return tmp.substr(1,tmp.length-2);
    if (tmp.length>5 && tmp.substr(0,4)=='<svg') return tmp;
     if (UI.usemathjax){
       tmp=giaceval('latex(quote('+tmp+'))');
       var dollar=String.fromCharCode(36);
       tmp=dollar+dollar+tmp.substr(1,tmp.length-2)+dollar+dollar;
       return tmp;
     }
     tmp=giaceval('mathml(quote('+tmp+',1))');
     tmp=tmp.substr(1,tmp.length-2);
    return tmp;   
  },
  ckenter:function(event,field,mode){
    var key = event.keyCode;
    if (key != 13 || event.shiftKey) return true;
   if (mode==3){ UI.quick(field.nextSibling); event.preventDefault(); field.select(); return true; }
    var tmp=field.value;
   Module.print(tmp);
    tmp=UI.caseval(tmp);
    if (mode==1){
      tmp=UI.rmquote(tmp); 
   }
   if (mode==2){
     tmp=UI.latexeval(tmp);
   }
   field.nextSibling.nextSibling.innerHTML=tmp;
   UI.render_canvas(field.nextSibling.nextSibling);
   if (UI.usemathjax) MathJax.Hub.Queue(["Typeset",MathJax.Hub,field.nextSibling.nextSibling]);
   if (event.preventDefault) event.preventDefault();
    return false;
  },
   exec: function(field){
     if (field.nodeName=="BUTTON" && field.innerHTML!='Clone' && field.innerHTML!='Restart' && field.innerHTML!='Exec. all'){
        field.click();
        return;
     }
     if (field.nodeName=="FORM"){
        UI.eval_form(field);
        return;
     }
     var f=field.firstChild;
     while (f){
       UI.exec(f);
       f=f.nextSibling;
     }
   },
  execonload: function(field){
     var f=field.nextSibling;
     if (f && f.innerHTML=="onload" && field.nodeName=="BUTTON"){
        field.click();
        return;
     }
     f=field.firstChild;
     while (f){
       UI.execonload(f);
       f=f.nextSibling;
     }
   },
  textarealtgt: function(field){
     if (field.nodeName=="TEXTAREA"){
        var tmp=UI.ltgt(field.value);
        field.value=tmp[0];
        //field.style.height=20*(tmp[1]+1)+'px';
        field.rows=tmp[1]+1;
        return;
     }
     var f=field.firstChild;
     while (f){
       UI.textarealtgt(f);
       f=f.nextSibling;
     }
   }
 };
 window.onload = function(e){
   var isFirefox = typeof InstallTrigger !== 'undefined';   // Firefox 1.0+
   var isSafari = Object.prototype.toString.call(window.HTMLElement).indexOf('Constructor') > 0;
  var ua = window.navigator.userAgent;
  var old_ie = ua.indexOf('MSIE ');
  var new_ie = ua.indexOf('Trident/');
  if ((!isFirefox && !isSafari) || (old_ie > -1) || (new_ie > -1) || window.chrome){
     UI.usemathjax=true;
     alert("Your browser does not support MathML, using MathJax for 2d rendering. Consider switching to Firefox for better rendering and faster results");
  }
  if (UI.usemathjax){
    var script = document.createElement("script");
    script.type = "text/javascript";
    script.src  = "https://cdnjs.cloudflare.com/ajax/libs/mathjax/2.7.0/MathJax.js?config=TeX-AMS-MML_HTMLorMML";
    document.getElementsByTagName("head")[0].appendChild(script);
  }
  var elem= document.getElementById('controlindex');
  elem.innerHTML='<hr><a href="'+String.fromCharCode(35)+'sec1">Table</a>, <a href="'+String.fromCharCode(35)+'sec2">Index</a>,'+elem.innerHTML;
  if (elem.style.maxHeight=='500px')
    elem.style.maxHeight=(window.innerHeight-120)+'px';
  elem= document.getElementById('maindiv');
  elem.style.maxHeight=(window.innerHeight-150)+'px';
  // Module.print(elem.innerHTML);
  //elem.parentNode.insertBefore(elem,null);
  giaceval=Module.cwrap('caseval',  'string', ['string']);
  giac3d = Module.cwrap('_ZN4giac13giac_rendererEPKc','number', ['string']);
  giaceval('set_language(1);');
  giaceval('factor(x^4-1)');
  giaceval('sin(x+y)+f(t)');
  UI.textarealtgt(document.documentElement);
  UI.execonload(document.documentElement);
  document.getElementById('output').innerHTML='';
 // if (confirm('Exec commands?')) UI.exec(document.documentElement);
 };
</script>
}
}
\else
\newcommand{\loadgiacscriptend}{}
\fi
\ifhevea
\newenvironment{giacprog}{
\verbatim}
{\endverbatim 
\@print{<button onclick="var field=parentNode.previousSibling; var tmp=field.innerHTML;if(tmp.length==0) tmp=field.value;var t=createElement('TEXTAREA');t.style.fontSize=16;t.cols=60;t.rows=10;var tmp1=UI.ltgt(tmp);t.value=tmp1[0];tmp=UI.caseval(tmp);tmp=UI.rmquote(tmp);nextSibling.innerHTML=tmp; UI.render_canvas(nextSibling.innerHTML); field.parentNode.insertBefore(t,field);field.parentNode.removeChild(field);">ok</button><span></span><br>
}
}
\newenvironment{giaconload}{
\verbatim}
{\endverbatim 
\@print{<button onclick="var field=parentNode.previousSibling; var tmp=field.innerHTML;if(tmp.length==0) tmp=field.value;var t=createElement('TEXTAREA');t.style.fontSize=20;t.cols=60;t.rows=UI.count_newline(tmp);var tmp1=UI.ltgt(tmp)[0];t.value=tmp1;tmp=UI.caseval(tmp);tmp=UI.rmquote(tmp);nextSibling.innerHTML=tmp; UI.render_canvas(nextSibling.innerHTML); field.parentNode.insertBefore(t,field);field.parentNode.removeChild(field);">ok</button><span>onload</span><br>
}
}
\else
\newenvironment{giacprog}
{
\VerbatimEnvironment
\begin{Verbatim}
}
{
\end{Verbatim}
}
\newenvironment{giaconload}
{
\VerbatimEnvironment
\begin{Verbatim}
}
{
\end{Verbatim}
}
\fi

\newcommand{\giaccmd}[3][style="width:400px;font-size:large"]{
\ifhevea
\@print{<textarea}
\@getprint{#1>#3}
\@print{</textarea><button onclick="var tmp=UI.caseval(}
\@getprint{'#2('}
\@print{+previousSibling.value+')');tmp=UI.rmquote(tmp); nextSibling.innerHTML='&nbsp;'+tmp;UI.render_canvas(nextSibling)">}
\@getprint{#2}
\@print{</button><span></span><br>}
\else
\lstinline@#2(#3)@
\fi
}
\newcommand{\giacinput}[2][style="width:400px;font-size:large"]{
\ifhevea
\@print{<textarea onkeypress="UI.ckenter(event,this,1)" }
\@getprint{#1>#2}
\@print{</textarea><button onclick="previousSibling.style.display='inherit';var tmp=UI.caseval(previousSibling.value);tmp=UI.rmquote(tmp); nextSibling.innerHTML='&nbsp;'+tmp;UI.render_canvas(nextSibling);">ok</button><span></span><br>}
\else
\lstinline@#2@
\fi
}
\newcommand{\giachidden}[3][style="display:none;width:400px;font-size:large"]{
\ifhevea
\@print{<textarea onkeypress="UI.ckenter(event,this,1)" }
\@getprint{#1>#2}
\@print{</textarea><button onclick="previousSibling.style.display='inherit';var tmp=UI.caseval(previousSibling.value);tmp=UI.rmquote(tmp); nextSibling.innerHTML='&nbsp;'+tmp;UI.render_canvas(nextSibling);">}
\@getprint{#3}
\@print{</button><span></span><br>}
\else
\lstinline@#2@
\fi
}
\newcommand{\giacinputbig}[2][style="width:800px;font-size:large"]{
\ifhevea
\@print{<textarea onkeypress="UI.ckenter(event,this,1)" }
\@getprint{#1>#2}
\@print{</textarea><button onclick="previousSibling.style.display='inherit';var tmp=UI.caseval(previousSibling.value);tmp=UI.rmquote(tmp); nextSibling.innerHTML='&nbsp;'+tmp;UI.render_canvas(nextSibling);">ok</button><span></span><br>}
\else
\lstinline@#2@
\fi
}
\newcommand{\giaccmdmath}[3][style="width:400px;font-size:large"]{\ifhevea
\begin{rawhtml}<br><textarea \end{rawhtml}
\@getprint{#1>#3}
\begin{rawhtml}</textarea><button onclick="var tmp=UI.caseval(\end{rawhtml} 
\@getprint{'#2('+}
\begin{rawhtml}previousSibling.value+')');  tmp=UI.latexeval(tmp);nextSibling.innerHTML='&nbsp;'+tmp; if (UI.usemathjax) MathJax.Hub.Queue(['Typeset',MathJax.Hub,nextSibling]);
">\end{rawhtml}
\@getprint{#2}
\begin{rawhtml}</button><span></span><br>\end{rawhtml}
\else
\lstinline@#2(#3)@
\fi
}
\newcommand{\giacinputmath}[2][style="width:400px;font-size:large"]{\ifhevea
\begin{rawhtml}<br><textarea onkeypress="UI.ckenter(event,this,2)" \end{rawhtml}
\@getprint{#1>#2} 
\begin{rawhtml}</textarea><button onclick="previousSibling.style.display='inherit';var tmp=UI.caseval(previousSibling.value); tmp=UI.latexeval(tmp);nextSibling.innerHTML='&nbsp;'+tmp; if (UI.usemathjax) MathJax.Hub.Queue(['Typeset',MathJax.Hub,nextSibling])">ok</button><span></span><br>\end{rawhtml}
\else
\lstinline@#2@
\fi
}
\newcommand{\giachiddenmath}[3][style="display:none;width:400px;font-size:large"]{\ifhevea
\begin{rawhtml}<br><textarea onkeypress="UI.ckenter(event,this,2)" \end{rawhtml}
\@getprint{#1>#2} 
\begin{rawhtml}</textarea><button onclick="previousSibling.style.display='inherit';var tmp=UI.caseval(previousSibling.value); tmp=UI.latexeval(tmp);nextSibling.innerHTML='&nbsp;'+tmp; if (UI.usemathjax) MathJax.Hub.Queue(['Typeset',MathJax.Hub,nextSibling])">\end{rawhtml}
\@getprint{#3}
\begin{rawhtml}</button><span></span><br>\end{rawhtml}
\else
\lstinline@#2@
\fi
}
\newcommand{\giaccmdbigmath}[3][style="width:800px;font-size:large"]{\ifhevea
\begin{rawhtml}<br><textarea \end{rawhtml}
\@getprint{#1>#3} 
\begin{rawhtml}</textarea><button onclick="var tmp=UI.caseval(\end{rawhtml} 
\@getprint{'#2('+}
\begin{rawhtml}previousSibling.value+')');  tmp=UI.latexeval(tmp);nextSibling.innerHTML=tmp; if (UI.usemathjax) MathJax.Hub.Queue(['Typeset',MathJax.Hub,nextSibling])">\end{rawhtml}
\@getprint{#2}
\begin{rawhtml}</button><div style="width:800px;max-height:200px;overflow:auto;color:blue;text-align:center"></div><br>\end{rawhtml}
\else
\lstinline@#2(#3)@
\fi
}
\newcommand{\giacinputbigmath}[2][style="width:800px;font-size:large"]{\ifhevea
\begin{rawhtml}<div><textarea onkeypress="UI.ckenter(event,this,2)" \end{rawhtml}
\@getprint{#1>#2} 
\begin{rawhtml}</textarea><button onclick="previousSibling.style.display='inherit';var tmp=UI.caseval(previousSibling.value); tmp=UI.latexeval(tmp);nextSibling.innerHTML=tmp; if (UI.usemathjax) MathJax.Hub.Queue(['Typeset',MathJax.Hub,nextSibling])">ok</button><div style="width:800px;max-height:200px;overflow:auto;color:blue;text-align:center"></div></div>\end{rawhtml}
\else
\lstinline@#2@
\fi
}
\newcommand{\giaclink}[2][Test online]{\ifhevea
\begin{rawhtml}<a href=\end{rawhtml}\@getprint{"#2"}
\begin{rawhtml} target="_blank">\end{rawhtml}
\@getprint{#1}
\begin{rawhtml}</a>\end{rawhtml}
\else
\fi
}
\newcommand{\giacslider}[6]{
\ifhevea
\begin{rawhtml}
<div><form onsubmit="setTimeout(function(){UI.eval_form(form);});return false;">
<input type="text" name="name" size="1" value=
\end{rawhtml}
\@getprint{"#1">}
\begin{rawhtml}
=<input type="number" name="valname" onchange="UI.eval_form(form);" value=
\end{rawhtml}
\@getprint{"#5">}
\begin{rawhtml}
<input type="button" value="-" onclick="valname.value -= stepname.value;UI.eval_form(form);">
<input type="button" value="+" onclick="valname.value -= -stepname.value;UI.eval_form(form);">
<input type="number" name="stepname" value=
\end{rawhtml}
\@getprint{"#4">}
\begin{rawhtml}
<input type="range" name="rangename"
onclick="valname.value=value;UI.eval_form(form);" value=
\end{rawhtml}
\@getprint{"#5" min="#2" max="#3" step="#4">}
\begin{rawhtml}
<textarea name="prog" onchange="UI.eval_form(form)" style="width:400px;vertical-align:bottom;font-size:large">
\end{rawhtml}\@getprint{#6}
\begin{rawhtml}
</textarea>
</form>
<span>Not evaled</span></div>
\end{rawhtml}
\else
\fi
}

%% file: castex.bbl
\begin{thebibliography}{10}

\bibitem{heveamathjax}
Yannick Chevallier.
\newblock {Hevea: LaTeX to HTML5 compiler, fork for MathJax support}.
\newblock {\tt https://github.com/YannickChevalier/hevea-mathjax}, 2017.

\bibitem{sagetex}
{Dan Drake}.
\newblock {SageTex}.
\newblock {\tt https://www.ctan.org/pkg/sagetex}, 2009.

\bibitem{itex2mml}
Jacques Distler.
\newblock {LaTeX to MathML converter}.
\newblock {\tt golem.ph.utexas.edu/\~\,distler/blog/itex2MML.html}, 2016.

\bibitem{pgiac}
{Jean-Michel Sarlat}.
\newblock {pgiac}.
\newblock {\tt http://melusine.eu.org/syracuse/giac/pgiac/}, 2011.

\bibitem{texmacs}
{Joris van der Hoeven}.
\newblock {Texmacs}.
\newblock {\tt http://www.texmacs.org/}, 2017.

\bibitem{hevea}
Luc Maranget.
\newblock {Hevea: LaTeX to HTML5 compiler (unstable version)}.
\newblock {\tt http://hevea.inria.fr/distri/unstable/}, 2017.

\bibitem{lyx}
{Matthias Ettrich}.
\newblock {Lyx}.
\newblock {\tt https://ww.lyx.org}, 2012.

\bibitem{jupyter}
{NumFOCUS Foundation}.
\newblock {The Jupyter Notebook}.
\newblock {\tt jupyter.org}, 2017.

\bibitem{hevea2mml}
Bernard Parisse.
\newblock {LaTeX to Mathml converter, fork for hevea output support}.
\newblock {\tt www-fourier.ujf-grenoble.fr/\~\,parisse/giac/heveatomml.tgz},
  2017.

\bibitem{giac}
Bernard Parisse and Ren\'ee~De Graeve.
\newblock {Giac/Xcas Computer Algebra System}.
\newblock {\tt www-fourier.ujf-grenoble.fr/\~\,parisse/giac.html}, 2017.

\bibitem{emscripten}
Alon Zakai.
\newblock {Emscripten: A C/C++ to Javascript compiler}.
\newblock {\tt kripken.github.io/emscripten-site/}, 2017.

\end{thebibliography}
